\documentclass[smallextended]{svjour3}       
\smartqed  
\usepackage{amsfonts}
\usepackage{lscape}
\usepackage{graphicx}
\usepackage{amsmath, mathtools}
\usepackage[clock]{ifsym}
\usepackage[ruled, vlined, linesnumbered, noresetcount]{algorithm2e}
\usepackage[square,numbers]{natbib}
\usepackage{multirow,booktabs}
\usepackage{array}
\usepackage{makecell}
\usepackage{rotating,url}
\usepackage{adjustbox}
\usepackage{rotating}

\SetKwRepeat{Do}{do}{while}

\usepackage{adjustbox}
\usepackage{longtable}
\usepackage[dvipsnames]{xcolor}
\usepackage{soul}

\title{Exact Methods for the Longest Induced Cycle Problem}
\author{Ahmad T.~Anaqreh \and Bogl\'arka G.-T\'oth \and Tam\'as Vink\'o}
\institute{Institute of Informatics, University of Szeged, Hungary \\ \email \{ahmad, boglarka, tvinko\}@inf.u-szeged.hu}
\date{draft version: \today}

\journalname{}
\begin{document}

\maketitle

\begin{abstract}

The longest induced (or chordless) cycle problem is a graph problem classified as NP-complete and involves the task of determining the largest possible subset of vertices within a graph in such a way that the induced subgraph forms a cycle. Within this paper, we present three integer linear programs specifically formulated to yield optimal solutions for this problem. The branch-and-cut algorithm has been used for two models. To demonstrate the computational efficiency of these methods, we utilize them on a range of real-world graphs as well as random graphs. Additionally, we conduct a comparative analysis against approaches previously proposed in the literature.

\end{abstract}

\section{Introduction}
\label{sec:intr}
\noindent
{\bf Cycles in graphs.} 
A significant part of combinatorial optimization is closely related to graphs. Within graph theory, the concept of graph cycles has fundamental importance. Identifying a simple cycle or a cycle with a specific structure within a graph forms the basis for numerous graph-theoretical problems that have been under investigation for many years. One such problem is the Eulerian walk, a cyclic path that traverses each edge exactly once, as discussed in \cite{west}. Another example is the Hamiltonian cycle, which traverses every vertex exactly once, as explored in \cite{akiyama}. 

\medskip
\noindent
{\bf Longest cycle.}
Kumar et al. \cite{kumar}, introduced a heuristic algorithm for the longest simple cycle problem. The authors utilized both adjacency matrices and adjacency lists, achieving a time complexity 
for the proposed algorithm
proportional to the number of nodes plus the number of edges of the graph. 
In \cite{alon}, the authors investigated the longest cycle within a graph with a large minimal degree. For a graph $G = (V, E)$ with a vertex count of $|V| = n$, the parameter $\mathrm{min\_deg}(G)$ denotes the smallest degree among all vertices in $G$, while ${\mathrm c}(G)$ represents the size of the longest cycle within $G$. The authors demonstrated that for $n > k \geq 2$, with  $\mathrm{min\_deg}(G) \geq n/k$, the lower bound ${\mathrm c}(G) \geq [n/(k - 1)]$ holds. Broersma et al.~\cite{broersma} proposed exact algorithms for identifying the longest cycles in claw-free graphs. A claw, in this context, refers to a star graph including three edges. The authors introduced two algorithms for identifying the longest cycle within such graphs containing $n$ vertices: one algorithm operates in $\mathcal{O}(1.6818^n)$ time with exponential space complexity, while the second algorithm functions in $\mathcal{O}(1.8878^n)$ time with polynomial space complexity.

\medskip 
\noindent
{\bf Longest isometric cycle.} 
In the work by Lokshtanov \cite{Lokshtanov}, the focus lies on the examination of the longest isometric cycle within a graph, which is defined as the longest cycle where the distance between any two vertices on the cycle remains consistent with their distances in the original graph. The author introduced a polynomial-time algorithm to address this specific problem.

\medskip
\noindent
{\bf Longest induced cycle.}
Our primary focus in this paper is dedicated to addressing the challenge of identifying the longest induced (or chordless) cycle problem. For a graph $G=(V, E)$ and a subset $W \subseteq V$, the $W$-induced graph $G[W]$ comprises all the vertices from set $W$ and the edges from $G$ that connect vertices exclusively within $W$. The objective of the longest induced cycle problem is to determine the largest possible subset $W$ for which the graph $G[W]$ forms a cycle. While it may seem straightforward to obtain an induced cycle since every isometric cycle is an induced cycle, it has been shown that identifying the longest induced cycle within a graph is an NP-complete problem, as demonstrated by Garey et al.~\cite{garey}.

The longest induced path ($P$), discussed in \cite{Matsypura}, represents a sequence of vertices within graph $G$, where each consecutive pair of vertices is connected by an edge $e \in E$ and there is no edge between non-consecutive vertices within $P$. In the context of a general graph $G$, determining the existence of an induced path with a specific length is proven to be NP-complete, as detailed in \cite{garey}. Consequently, the longest induced cycle can be considered as a special case of the longest induced path.

Holes in a graph, defined as induced cycles with four or more vertices, play a significant role in various contexts. Perfect graphs, for instance, are characterized by the absence of odd holes or their complements \cite{chen2007}. Moreover, when addressing challenges like finding maximum independent sets in a graph \cite{nemhauser1992}, the existence of odd holes leads to the formulation of odd hole inequalities, strengthening approaches for these problems. Similarly, in other problem domains such as set packing and set partitioning \cite{borndorfer1998}, these odd hole inequalities serve as crucial components.

Several papers have explored the longest induced cycle problem in graphs with specific structures. In \cite{fuchs}, the author investigated the longest induced cycle within the unit circulant graph. To define the unit circulant graph $X_n=Cay(\mathbb{Z}_n;\mathbb{Z}_n^*)$, where $n$ is a positive integer, consider the following. The vertex set of $X_n$, denoted as $V(n)$, comprises the elements of $\mathbb{Z}_n$, the ring of integers modulo $n$. The edge set of $X_n$, represented as $E(n)$, for $x,y \in V(n)$, $(x,y) \in E(n)$ if and only if $x - y \in \mathbb{Z}_n^*$, with $\mathbb{Z}_n^*$ being the set of units within the ring $\mathbb{Z}_n$. The author demonstrates that if the positive integer $n$ has $r$ distinct prime divisors, then $X_n$ contains an induced cycle of length $2^r+2$.
In a separate study by Wojciechowski et al.~\cite{wojciechowski}, the authors examine the longest induced cycles within hypercube graphs. If $G$ represents a $d$-dimensional hypercube, they proved the existence of an induced cycle with a length $\geq (9/64)\cdot 2^d$.

Almost parallel to our work, Pereira et al. \cite{pereira} dealt with the longest cordless cycle problem, which is equivalent to the longest induced cycle problem. They presented an integer linear program (ILP) formulation along with additional valid inequalities to strengthen and refine the formulation, all of which were incorporated into a branch-and-cut algorithm. They applied a multi-start heuristic method for initial solution generation and then conducted performance evaluations of the algorithm on a range of randomly generated graphs, including those with up to 100 vertices. They could solve the largest problems within 3,011.17 seconds. Our aim is to provide models and methods that can work more efficiently. The models and the best branch-and-cut versions proposed in \cite{pereira} are discussed in Section \ref{Cordless Cycle formulation}.

Our paper proposes three integer linear programs (ILPs) designed to handle the longest induced cycle problem within general graphs. Some of these models were built based on those used in previous work focused on solving the longest induced path problem, as seen in the studies by Marzo et al.~\cite{Marzo} and Bokler et al.~\cite{Bokler}. Matsypura et al.~\cite{Matsypura} introduced three integer programming (IP) formulations and an exact iterative algorithm based on these IP formulations for tackling the longest induced path problem. However, it is important to note that we do not extend these methods, as they were found to be less effective compared to models in \cite{Marzo, Bokler}.

The rest of the paper is organized as follows. Section \ref{sec:meth} and \ref{sec:Algorithms} discuss the models and methodologies used to solve the problem together with the best models and methods presented in \cite{pereira}. Section \ref{sec:numexp} reports the numerical results to show the efficiency of our models, also compared to the results in \cite{pereira}. The conclusion of our work is presented in Section \ref{sec:con}.

\section{Models}
\label{sec:meth}
\subsection{Notations}

Let $G=(V, E)$ be an undirected graph with vertex set $V$ and edge set $E\subset V\times V$. An edge $e\in E$ can be given as $(i,j)$ for some $i,j\in V$. However, the symmetric pair $\Bar{e}=(j,i)$ is not included in $E$. Thus, we introduce the symmetric edge set $E^*= E\cup \{\Bar{e}=(j,i): e=(i,j) \in E\}$. Throughout, an edge $e=(i,j)$, and $\bar{e}=(j,i)$ unless explicitly stated otherwise.

We use the notation $\delta$ for adjacent edges over vertices and edges as follows. Let us denote the outgoing and incoming edges incident to vertex $i$ with $\delta^+(i)=\{(i,k) \in E^*\}$, and $\delta^-(i)=\{(k,i) \in E^*\}$, respectively. Additionally, $\delta(i)=\delta^+(i) \cup \delta^-(i)$ denote all the edges incident to vertex $i$.

For an edge $e=(i,j)\in E^*$, outgoing edges are $\delta^+(e)=\delta^+(i) \cup \delta^+(j)\setminus \{e,\bar{e}\}$, and similarly, incoming edges are $\delta^-(e)=\delta^-(i) \cup \delta^-(j)\setminus \{e,\bar{e}\}$. The neighbour edges of $e$ are denoted by $\delta(e)=\delta^+(e) \cup \delta^-(e)$ for all $e\in E^*$.
This notation can be extended to any subset of vertices $C\subset V$, where $\delta^+(C):=\{(k,l) \in E^*: k \in C ,l \in V\setminus C\}$ and $\delta^-(C):=\{(k,l) \in E^*: l \in C ,k \in V\setminus C\}$ denote the outgoing and incoming edges of $C$, respectively, and $\delta(C)= \delta^+(C) \cup \delta^-(C)$ all edges that connect $C$ with $V\setminus C$.



\subsection{Order-based model}
\label{sec:LIC}
The first model to discuss, called \emph{LIC}, is an MILP model using order-based formulation to avoid subtours. The formalism of the model is as follows:
\begin{equation}
\max  \sum_{i\in V}y_i
\label{lic:obj}
\end{equation}
subject to
\begin{alignat}{10}
x_{e}+x_{\Bar{e}} &\leq 1 &&\quad\forall\ e\in E& \label{lic:c1}\\
\sum_{g \in \delta^+(e)}x_{g}&\leq 1&&\quad\forall\ e\in E^*&\label{lic:c2}\\
y_i &= \sum_{\substack{g \in \delta^+(i)}}x_{g} &&\quad\forall i\in V\label{lic:c3}\\
y_i &= \sum_{\substack{g \in \delta^-(i)}}x_{g} &&\quad\forall i\in V&\label{lic:c4}\\
\sum_{i\in V}w_{i} &= 1 &\label{lic:c5}\\
w_{i} &\leq y_{i}  &&\quad\forall\ i \in V& \label{lic:c6}\\
u_i-u_j &\leq n(1-x_{e})-1+n w_{i} &&\quad \forall e\in E^*&\label{lic:c7}\\
\sum_{i\in V}i w_{i} &\leq j y_{j}+n(1-y_{j}) &&\quad \forall\ j \in V&\label{lic:c8}\\
&y_{i}, u_i \geq  0 &&\quad\forall i \in V\\
&x_{e} \in \{0,1\} &&\quad\forall e \in E^*\\
&w_{i} \in \{0,1\} &&\quad\forall i \in V 
\end{alignat}

The variable $y_{i}$ indicates whether vertex $i$ is part of the longest induced cycle or not. Consequently, the objective in \eqref{lic:obj} aims to maximize the sum of these variables, which directly corresponds to the length of the cycle. The decision variable $x_{e}$ is one if the edge $e$ is included in the solution, and zero otherwise.

The constraints can be understood as follows. Given that $E^*$ is symmetric, constraint \eqref{lic:c1} guarantees that only one of the edges $e$ or $\Bar{e}$ can exist in the cycle, preventing the formation of small cycles. Constraint \eqref{lic:c2} ensures that for any edge $e=(i,j)\in E^*$, only one outgoing edge from either vertex $i$ or vertex $j$ can be part of the cycle. Constraints \eqref{lic:c3} and \eqref{lic:c4} ensure that for a given vertex $i$, only one outgoing edge and one incoming edge can be chosen to be part of the cycle. The constraint \eqref{lic:c7} is a modified Miller-Tucker-Zemlin (MTZ) order-based formulation: if edge $e$ is in the cycle, vertices $i$ and $j$ must be arranged in sequential order unless the binary variable $w_i$ equals 1. This variable is introduced to handle the position of the last vertex in the cycle, facilitating the ordering process. Constraint \eqref{lic:c8} functions as a symmetry-breaking constraint, as described in \cite{walsh}. It enforces that the last vertex in the cycle must have the smallest index among all vertices in the cycle.

\medskip
For a variation of the above introduced \emph{LIC} model consider the following constraint:
\begin{alignat}{10}
x_{e}+x_{\Bar{e}} &\geq y_i+y_j-1 &&\qquad\forall\ e=(i,j)\in E
\label{lic:c11}
\end{alignat}
Constraint \eqref{lic:c11} guarantees that either edge $e$ or $\Bar{e}$ must be included in the solution if both endpoints $i$ and $j$ are part of the solution. Conversely, if an edge is not selected for the solution, neither of its endpoints can be included in the solution. By substituting constraint \eqref{lic:c2} in the original \emph{LIC} model with constraint \eqref{lic:c11}, we create a new model, \emph{LIC}2. This modification leads to improved runtime performance compared to \emph{LIC}, as demonstrated in Section \ref{sec:numexp}.

\subsection{Subtour-elimimation model}
The second model we employ to address the longest induced cycle problem is based on the model presented by Bokler et al.~\cite{Bokler}, which is referred to \emph{ILP}$_{cut}$ and was originally designed for identifying the longest induced path. $E^*$ is the symmetric edge set, as defined previously. Let $\mathcal{C}$ represent the set of cycles in $G$. The model is defined as follows: 
\begin{equation}
\max \frac12 \sum_{e\in E^*}x_{e} 
\label{ilp:obj}
\end{equation}
subject to
\begin{alignat}{10}
x_{e} &= x_{\Bar{e}}&&\qquad \forall e \in E& \label{ilp:c1}\\
x_{e} &\leq \sum_{g \in \delta^-(i)} x_{g} &&\qquad \forall\ e=(i,j) \in E^*& \label{ilp:c2}\\
\sum_{g \in \delta^-(i)} x_{g} + \sum_{g \in \delta^+(j)} x_{g} & \leq 2 &&\qquad \forall\ e=(i,j) \in E^*& \label{ilp:c3}\\
\sum_{e \in \delta(i)} x_{e} &\leq \sum_{g \in \delta(C)} x_{g} &&\qquad \forall C \in \mathcal{C}, i \in C& \label{ilp:c4}\\
&x_{e} \in \{0,1\} &&\qquad \forall e\in E^* 
\end{alignat}

The binary decision variable $x_{e}$ indicates whether edge $e$ is a part of the longest induced cycle, but unlike in the \emph{LIC} model (in Section \ref{sec:LIC}), in this case, edge selection is symmetric. Consequently, the objective is to maximize half of the sum of these variables, as defined in objective function \eqref{ilp:obj}. Symmetry of the solution is guaranteed by \eqref{ilp:c1}. Constraint \eqref{ilp:c2} enforces that the solution forms a cycle or cycles, while constraint \eqref{ilp:c3} specifies that for any edge $e$, precisely two of its adjacent edges must also be selected. This ensures the induced property of the solution. Constraint \eqref{ilp:c4} is utilized to eliminate small cycles in the graph.

\subsection{Cycle-elimination model}
Our third model, called \emph{cec}, is a modified version of the \emph{cec} model introduced in \cite{Marzo} to find the longest induced path. In this model, the symmetry of the edges is not used. The formalism of the model is as follows:
\begin{equation}
\max  \sum_{i\in V}y_i
\label{$cec$:obj}
\end{equation}
subject to
\begin{alignat}{10}
\sum_{e \in \delta(i)} x_{e} &= 2y_{i}&&\quad \forall\ i \in V& \label{$cec$:c1}\\
x_{e}&\leq y_{i}&&\quad \forall\ i \in V,e \in \delta(i)& \label{$cec$:c2}\\
x_{e}&\geq y_{i}+y_{j}-1&&\quad \forall e=(i,j) \in E& \label{$cec$:c3}\\
\sum_{i \in C}y_{i}&\leq |C|-1 &&\quad \forall C \in \mathcal{C}& \label{$cec$:c4}\\
&y_{i} \in \{0,1\} &&\quad  \forall i \in V\\
&x_{e} \in \{0,1\} &&\quad \forall e \in E 
\end{alignat}

The binary decision variable $y_i$ maintains its previous interpretation, equal to one if vertex $i$ is part of the solution. Additionally, variable $x_{e}$ is set to one if edge $e$ is included in the solution. However, in this context, the symmetric edge is not needed. The objective function \eqref{$cec$:obj} seeks to maximize the number of vertices within the induced cycle. Constraint \eqref{$cec$:c1} guarantees that each vertex within the solution is connected to precisely two vertices in the cycle. Constraints \eqref{$cec$:c2} and \eqref{$cec$:c3} are in place to ensure that the cycle is induced. To eliminate solutions composed of small cycles from consideration, constraint \eqref{$cec$:c4} is introduced. $\mathcal{C}$ represents a set of the cycles for the given graph. Constraint \eqref{$cec$:c4} is combined into the model to enforce the solution to consist of a single cycle. This means that multiple small cycles are not deemed valid solutions.

\subsection{Cordless-cycle model}
\label{Cordless Cycle formulation}
The CCP formulation was introduced by Pereira et al.~\cite{pereira} to deal with the problem at hand. 
%
The CCP formulation is formally described as follows:

\begin{equation}
\max \sum_{i \in V} y_i
\label{BC:obj}
\end{equation}
subject to
\begin{alignat}{10}
\sum_{e \in E} x_e &= \sum_{i \in V} y_i& \label{BC:c1}\\
\sum_{i \in V} y_i &\geq 4& \label{BC:c2}\\
\sum_{e \in \delta(i)} x_e &= 2y_i&&\quad \forall\ i \in V& \label{BC:c3}\\
\sum_{g \in \delta(C)} x_g&\geq 2(y_i+y_j-1)&&\quad C \subset V, i \in C,j \in V\setminus C& \label{BC:c4}\\
x_{e}&\leq y_{i}&&\quad \forall\ i \in V,e \in \delta(i)& \label{BC:c5}\\
x_e &\geq y_i+y_j-1&&\quad \forall\ e=(i,j) \in E& \label{BC:c7}\\
&x_{e} \in \{0,1\} &&\quad \forall e \in E \label{BC:c8}\\
&y_{i} \in \{0,1\} &&\quad  \forall i \in V\label{BC:c9}
\end{alignat}
The formulation includes the usual sets of binary variables: $y_i$ and $x_e$, indicating whether vertex $i$ and edge $e$ are in the cycle or not.
Consequently, the number of selected vertices and edges is equal, as required by (\ref{BC:c1}), and at most four vertices must be selected by \eqref{BC:c2}. Each vertex within the solution is incident to precisely two edges, as guaranteed by (\ref{BC:c3}). Moreover, the subgraph defined by $x$ and $y$ remains connected, as guaranteed by the subtour elimination constraint (\ref{BC:c4}). Furthermore, \eqref{BC:c5}–\eqref{BC:c7} ensures that any solution is an induced subgraphs of $G$. More specifically, any edge of $G$ with both its endpoints belonging to the solution must be part of the solution. 

The CCP formulation was employed by the authors of \cite{pereira}, along with various valid inequalities. They introduced nine branch-and-cut (\emph{BC}) algorithms and subsequently chose the top three among them. The first one, labeled as \emph{BC}1 contains constraints \eqref{BC:c1}-\eqref{BC:c9}, and in addition the following constraint:
\begin{alignat}{10}
\sum_{g \in \delta(C)} x_g &\geq 2x_e&&\quad  C \subset V,e=(i,j) \in E,i \in C,j \in V\setminus C
\label{BC:c13}
\end{alignat}
This algorithm initiates by separating \eqref{BC:c4}, and subsequently, the resulting inequality is enhanced to the more robust form of \eqref{BC:c13}. This specific constraint ensures that if $x_e=1$, then it is mandatory for $y_i=y_j=1$ to hold true, due to the presence of inequalities \eqref{BC:c5}-\eqref{BC:c7}.

For the \emph{BC}2 and \emph{BC}3 algorithms, both constraints (\ref{BC:c14}) and (\ref{BC:c17}) were included together with constraints \eqref{BC:c1}-\eqref{BC:c13}. 
\begin{alignat}{10}
\sum_{i \in Q} y_i&\leq 2
\label{BC:c14}
\end{alignat}
\begin{alignat}{10}
\sum_{e \in E(Q)} x_e  &\geq \sum_{i \in Q} y_i-1
\label{BC:c17}
\end{alignat}
For a clique $Q \subset V$, $|Q| \geq 3$. Constraint (\ref{BC:c14}) ensures that within a clique $Q$ at most two of its vertices can be part of the induced cycle. On the other hand, constraint (\ref{BC:c17}) guarantees that for a clique $Q$ the number of vertices that can be part of the induced cycle is limited to at most one more than the number of edges that can be included from the clique. Namely, only one of the edges from $Q$ might be included in the solution.

For the \emph{BC}2 algorithm, they implemented a rule that imposes no restrictions on the number of separation rounds. In other words, whenever a violated inequality is detected, it is included in the cut pool. Conversely, for \emph{BC}3, a fixed number of separation rounds, specifically 30, was established, and inequalities were added to the cut pool if a clique did not share two or more vertices with a clique in a previously accepted inequality. The order of inequalities in the cut pool was determined by descending order of the absolute values of their corresponding linear programming relaxation dual variables. All three algorithms utilized the lower bounds obtained from the multi-start CCP heuristic, which is a constructive procedure that takes a predefined edge as input data. The algorithm then seeks to extend a tentative path, $P$, containing the selected vertices. Vertices are added to $P$ one at a time, accepted if they are adjacent to one of the path's current extremities and not adjacent to any internal vertices. The procedure terminates when the endpoints of $P$ meet, resulting in a chordless cycle of $G$, or when further expansion of $P$ becomes impossible.

\section{Algorithms}
\label{sec:Algorithms}
Out of the three models we have introduced, only \emph{LIC} (and \emph{LIC}2) can be directly solved using any MILP solver. Both $ILP_{cut}$ and $cec$ rely on the set of small cycles, which are usually created as part of the solution process, either through an iterative cut generation approach or, more effectively, via branch-and-cut algorithm by employing separation.


Note that subtour elimination inequalities \eqref{ilp:c4} and \eqref{$cec$:c4}, present in the \emph{ILP}$_{cut}$ and \emph{cec} models respectively, exhibit exponential complexity. Consequently, attempting to enumerate all inequalities corresponding to each subtour within the graph and subsequently cutting them becomes impractical. Instead, we have added these inequalities to the \emph{ILP}$_{cut}$ and \emph{cec} models as soon as facing them. Hence, the cut generation approach is employed as follows: the method is initiated with a model relaxing all subtour elimination inequalities, and if subtours arise in integer solutions, violated inequalities are added, 
and this process is repeated until the optimal solution is reached. For that, callback functionality from Gurobi \cite{gurobi} was employed, which can be used to add these inequalities iteratively.

We employed the Depth-First Search (DFS) algorithm on the induced subgraph of the integer solution to identify cycles, subsequently introducing a new inequality for each subtour discovered. 
The entire procedure, which combines the models and cut generations, is shown in Algorithms \ref{alg:CallBack1} and \ref{alg:CallBack2}. 

\subsection{Initialization}

In the initialization phase of the procedure, the \emph{ILP}$_{cut}$ and \emph{cec} models are created, encompassing the creation of their variables, constraints, and objective functions.


\subsection{Cut generation}

To tackle the \emph{ILP}$_{cut}$ and \emph{cec} models, we combine the cut generation mechanism with the Branch-and-bound method, as explained earlier in this Section. Consequently, each model was addressed using two distinct methodologies, as outlined below.

\subsubsection{Soft approach}
The first approach involves cut generation as outlined in Algorithm \ref{alg:CallBack1}. In each iteration, a subproblem from the Branch-and-Bound tree is solved. 
In line  \ref{$cec$feas} the algorithm checks if the solution of the subproblem is an integer solution. Based on this, the DFS algorithm is employed to detect any subtours within the solution, as shown in line \ref{$cec$cycle}. If a subtour exists, and its length is less than or equal to the value of the variable {\tt longest\_induced\_cycle}, a cut is appended for that cycle. If not, the value of the variable is updated to reflect the length of the cycle, and there is no need to introduce a cut because the cycle could potentially be the optimal solution. These details are clarified in lines \ref{ceccyclength} through \ref{cecupdatecylen}. The cut generation terminates when there are no further subtours present in the solution, indicating the completion of the procedure.

\begin{algorithm}
\label{alg:CallBack1}
\caption{Cut\_Generation1}
\SetKwInOut{Input}{Input}
\SetKwFunction{CECC}{Cut\_Generation1}
\SetKwFunction{CECInit}{model Initialization}
\SetKwProg{Fn}{Function}
\DontPrintSemicolon
\SetKwFunction{}{}
\CECInit{} \tcp*[f]{setting up the $ILP_{cut}$ or $cec$ model components}\\
{\tt longest\_induced\_cycle=0}\\
\Fn{\CECC{}\label{$cec$call}}
{
   \If(\tcp*[f]{model has integer solution} \label{$cec$feas}){\tt model.status$==$feasible\_integer}
      {
      { \tt C=DFS(feasible\_integer)} \tcp*[f]{find subtour in the solution}\label{$cec$cycle}\\
    \If(\label{ceccyclength})
        {\tt length(C) $\leq$ longest\_induced\_cycle}
        {
            {\tt model.addConstr(\ref{ilp:c4},\ref{$cec$:c4})} \tcp*[f]{add cut \eqref{ilp:c4} or \eqref{$cec$:c4} to the model}\label{$cec$cut1}\\
        }
         \Else
         {
            {\tt longest\_induced\_cycle=length(C)} \tcp*[f] {update variable value}\label{cecupdatecylen}
         }
      }
      
}
{ \tt model.optimize(\CECC{})} \tcp*[f]{solve the model using cut generation}\label{opt$cec$}\\
{ \tt print(longest\_induced\_cycle)}
\end{algorithm}

\subsubsection{Tough approach}

The second cut generation-based approach is detailed in Algorithm \ref{alg:CallBack2}. In each iteration, a subproblem is solved, and if an integer solution is obtained, the algorithm verifies the presence of any subtours using the DFS algorithm, as described in lines \ref{$cec$feas2} through \ref{$cec$cycle2}. If any cycles are detected, a cut is integrated into the model (line \ref{$cec$cut}), and the length of the cycle is updated if it exceeds the value of the variable {\tt longest\_induced\_cycle} (line \ref{$cec$cyclenupdate}). Although we may cut the optimal induced cycle, its length (and possibly the cycle itself) is recorded. It is important to note that  in order to further improve the procedure, a constraint is added to the model in line \ref{$cec$cons}. This constraint ensures that the objective value must be greater than the length of the largest induced cycle discovered so far. By using this cut generation, the branch-and-cut indicates the infeasibility of the problem, yet the longest induced cycle length recorded in the variable {\tt longest\_induced\_cycle}.
 \begin{algorithm}
\label{alg:CallBack2}
\caption{Cut\_Generation2}
\SetKwInOut{Input}{Input}
\SetKwFunction{CECCtwo}{Cut\_Generation2}
\SetKwFunction{CECInit}{model Initialization}
\SetKwProg{Fn}{Function}
\DontPrintSemicolon
\SetKwFunction{}{}
\CECInit{} \tcp*[f]{setting up the $ILP_{cut}$ or $cec$ model components}\\
{\tt longest\_induced\_cycle=0}\\
\Fn{\CECCtwo{}\label{$cec$call2}}
{
   \If(\tcp*[f]{model has integer solution} \label{$cec$feas2}){\tt model.status$==$feasible\_integer}
      {
      { \tt C=DFS(feasible\_integer)} \tcp*[f]{find subtour in the solution}\label{$cec$cycle2}\\
       {\tt model.addConstr(\ref{ilp:c4},\ref{$cec$:c4})} \tcp*[f]{add cut \eqref{ilp:c4} or \eqref{$cec$:c4} to the model}\label{$cec$cut}\\
      \If( \label{liclength2})
        {\tt length(C) $>$ longest\_induced\_cycle}
        {
            {\tt longest\_induced\_cycle=length(C)}\label{$cec$cyclenupdate}
         }
         {\tt model.addConstr(model.ObjVal $\geq$ longest\_induced\_cycle+1)} 
         \label{$cec$cons}\\
      }
}
{ \tt model.optimize(\CECCtwo{})} \tcp*[f]{solve the model using cut generation}\label{opt$cec$2}\\
{ \tt print(longest\_induced\_cycle)}
\end{algorithm}

\subsection{Longest Isometric Cycle}

Lokshtanov's algorithm, as described in \cite{Lokshtanov}, aims to identify the longest isometric cycle within a graph. In accordance with the definition of an isometric cycle, as discussed in Section \ref{sec:intr}, if a given graph $G$ contains an isometric cycle with a length of $\ell$, then there must also exist an induced cycle within the graph with a length of $m$ where $m \geq \ell$. Consequently, the longest isometric cycle serves as a benchmark for the longest induced cycle. The algorithm's objective is to verify the existence of an isometric cycle with a length of $k$ in a given graph $G=(V, E)$. If such a cycle exists, the graph $G$ can be employed to construct a new graph $G_k$ with vertices as vertex-pairs of $G$. Namely, $V(G_k)=\{(u,v) \in V: d(u,v)= \lfloor k/2 \rfloor\}$, where $d(u,v)$ is the length of the shortest path between $u$ and $v$, and its edge set given by $E(G_k)=\{((u,v),(w,x)): (u,w) \in E(G) \wedge (v,x) \in E(G)\}$.

\medskip

The method is outlined in Algorithm \ref{alg:LIC}. For a given value of $k$, the algorithm computes the graph $G_k$ and examines whether there exists a pair of vertices $(u,v)$ and $(v,x)$ within $V(G_k)$ such that $(v,x)$ belongs to the set $ M_k(u,v) := \{(u,x): (u,x) \in V(G_k) \wedge (v,x) \in E(G)\}$ and $d_{G_k}[(u,v),(v,x)]=\lfloor{k/2}\rfloor$. If such a pair is found, it indicates the presence of an isometric cycle with a length of $k$. 

\begin{algorithm}
\SetKwInOut{Input}{Input}
\SetKwProg{Fn}{Function}
\DontPrintSemicolon
\SetKwFunction{}{}
{\tt LISC=0}\\
\For(\tcp*[f]{distance calculation})
{\tt $\forall l \in V,i \in V, j \in V$\label{dist1}}
{\tt $d(i,j) = \min\{d(i,j),d(i,l)+d(j,l)\}$ \tcp*[f]{by Floyd algorithm}\label{diste2}}

\If(\tcp*[f]{no cycles in tree graph}){$G$ \tt is a tree}
      {
        {\tt return LISC}
      }
\For(){\tt $k = 3 \to n$}
    {
    $V_{k}=\emptyset$ \tcp*[f]{vertices of $G_k$}\\
    \For{\tt $u \wedge v \in V$}
        {
         \If{\tt $d(u,v)= \lfloor k/2 \rfloor$}
                {
                    {\tt $V_k=V_k \cup \{(u,v)\}$}
                }
        }
    $E_{k}=\emptyset$  \tcp*[f]{edges of $G_k$}\\
    \For{\tt $(u,v) \wedge (w,x) \in V_{k}$}
        {
         \If{\tt $(u,w) \in E \wedge (v,x) \in E $}
                {
                    {\tt $E_{k}=E_{k} \cup \{((u,v),(w,x))\}$}
                }
        }
    $G_{k}=(V_{k},E_{k})$\\
    \For{\tt $(u,v,x) \in V$}
        {
         \If{\tt $(u,v) \in V_{k} \wedge (v,x) \in M_{k}(v,u) \wedge d_{G_{k}}[(u,v),(v,x)] = \lfloor k/2 \rfloor  $}
                {
                    {\tt LISC$:=k$}
                }
        }
    }
\tt {print(LISC)}
\caption{Longest Isometric Cycle}\label{alg:LIC}
\end{algorithm}
\section{Numerical experiments}
\label{sec:numexp}

To demonstrate and evaluate the effectiveness of the proposed methods, we present numerical results for three models: the \emph{LIC} model, the \emph{ILP}$_{cut}$ model, 
and the \emph{cec} model. 
Furthermore, we conducted a comparison between our best results and results from \cite{pereira} on randomly generated graphs to highlight the efficiency of our approach in comparison to existing methods.

\subsection{Computational environment and datasets}

The algorithms detailed in Section \ref{sec:Algorithms} were implemented in Julia 1.7.0, utilizing the JuMP package version 0.22.1. We employed Gurobi 9.5.0 as the solver for all experiments. Each run was constrained to a one-hour time limit and a single thread. For the longest isometric cycle algorithm, we implemented it using Python 3.8 with a 24-hour time limit. These computations were performed on a computer with an Intel Core i7-4600U CPU, 8GB of RAM, and running the Windows 10 operating system.

To verify the efficacy of our methods, we employed two sets of network datasets. The first is the RWC set, comprising 19 real-world networks that encompass communication and social networks within companies, networks of book characters, as well as transportation, biological, and engineering networks, as described in \cite{Matsypura}. Additionally, we utilized the Movie Galaxy (MG) set, consisting of 773 graphs that represent social networks among movie characters, as detailed in \cite{kaminski}. For further information about these instances, the reader is referred to the following link: \url{http://tcs.uos.de/research/lip}.

To perform a comparison with the results presented in \cite{pereira}, we conducted experiments on random graphs with varying values of $n$ ranging from 50 to 100, considering both 10\% and 30\% density, as in \cite{pereira}. For every case, 10 graphs were generated. Every run was restricted to a maximum duration of one hour, with no restrictions on the number of threads, and with an initial solution set to 4, as described in \cite{pereira}. Regarding the hardware comparison, we utilized the information available in \cite{cpu} to collect the details of the CPU utilized in all experiments, as outlined in Table \ref{table:cpu}. It is evident that the computer used in \cite{pereira} is more powerful than ours. To ensure a fair comparison, we normalized the execution times in all cases. The ratio between the single-thread ratings gives a good approximation of the relative speed. Therefore, we calculated this ratio in the last row of Table \ref{table:cpu}. The run time was then modified by multiplying it by the obtained ratio.

\begin{table}[th!]
\centering
\caption{CPU performance comparison between the CPU used in this paper and in \cite{pereira}.}
\begin{tabular}{c|c|c}
\toprule
Benchmarks&\thead{Intel Core i7-4600U}&\thead{Intel Xeon W-3223 }\\
\midrule
Clock Speed (GHz)&2.1&3.5\\
Turbo Speed (GHz)&Up to 3.3&Up to 4.0 GHz\\
Number of Physical Cores&2 (Threads: 4)&8 (Threads: 16)\\
Single Thread Rating&1641&2480\\
\midrule
Ratio &0.66&1\\
\bottomrule
\end{tabular}
\label{table:cpu}
\end{table}

\subsection{Computational results}
Table \ref{table:exper1} presents the computational experiments conducted on the RWC instances. The second column displays the optimal solutions for each instance (opt). In the third column, we find the length of the longest isometric cycle (\emph{LISC}), if possible. The fourth and fifth columns respectively indicate the number of vertices ($N$) and edges ($M$) of the corresponding graph. Columns six through eleven show the time in seconds required to identify the optimal solution using the various methods employed in this study. Specifically, \emph{ILP}$_{cut}$2 and \emph{cec}2 refer to the methods outlined in Algorithm \ref{alg:CallBack2}. For all these methods, we initiated the search using the \emph{LISC} value, incorporating the constraint {\tt ObjVal $\geq$ LISC}. These methods are indicated in every second row corresponding to each graph. Instances that resulted in timeouts are denoted by the symbol {\small \showclock{1}{00}}.

\begin{center}
\small
\begin{longtable}{lr@{\ }c@{\ }r@{\ }r@{\ \ }r@{\ \ }r@{\ \ }r@{\ \ }rrr}
\caption{Running times on RWC instances, time is given in seconds.}\\
\toprule
graph & opt & LISC  & $N$ & $M$  & \emph{LIC} & \emph{LIC}2 & \emph{ILP$_{cut}$}  & \emph{ILP$_{cut}2$} & \emph{cec}    & \emph{cec}2 \\
\midrule\midrule\endhead 

\multirow{2}{*}{high-tech}& \multirow{2}{*}{10}  & \multirow{2}{*}{5}& \multirow{2}{*}{33}  & \multirow{2}{*}{91}  &1.22&0.63&0.95&0.33&0.38&0.14\\
&&&&&0.99&0.42&1.15&1.57&1.02&0.77\\
\midrule
\multirow{2}{*}{karate} & \multirow{2}{*}{6}& \multirow{2}{*}{5}& \multirow{2}{*}{34}& \multirow{2}{*}{78}  & 0.63&0.58&0.21&0.24&0.20&0.19  \\
&&& && 0.66&0.53&0.23&0.29&0.24&0.17 \\
\midrule
\multirow{2}{*}{mexican}& \multirow{2}{*}{13}  & \multirow{2}{*}{7}& \multirow{2}{*}{35}  & \multirow{2}{*}{117} & 0.92&0.82&0.66&0.88&0.24&0.20\\
&&&&&0.83&0.78&0.69&0.74&0.20&0.20\\
\midrule
\multirow{2}{*}{\underline{\textbf{sawmill}}} & \multirow{2}{*}{6}   & \multirow{2}{*}{5}   & \multirow{2}{*}{36}  & \multirow{2}{*}{62}  
&0.54&0.43&\underline{\textbf{0.18}}&0.37&\underline{\textbf{0.15}}&0.10\\
&&&&&0.33&0.30&\underline{\textbf{0.36}}&0.16&\underline{\textbf{0.13}}&0.11\\
\midrule
\multirow{2}{*}{tailorS1}       & \multirow{2}{*}{12}  & \multirow{2}{*}{7}   & \multirow{2}{*}{39}  & \multirow{2}{*}{158} &2.93&1.11&1.37&1.45&0.34&0.33\\
&&&& & 1.38&0.89&1.76&1.76&0.37&0.44 \\
\midrule
\multirow{2}{*}{chesapeake}     & \multirow{2}{*}{15}  & \multirow{2}{*}{5}   & \multirow{2}{*}{39}  & \multirow{2}{*}{170} & 1.01&0.69&0.56&0.81&0.24&0.22\\
&&&&&1.05&0.72&0.97&0.87&0.28&0.31\\
\midrule
\multirow{2}{*}{tailorS2}       & \multirow{2}{*}{12}  & \multirow{2}{*}{5}   & \multirow{2}{*}{39}  & \multirow{2}{*}{223} & 3.11&2.05&3.49&4.46&0.74&0.65\\
&&&& &3.25&3.09&3.74&4.62&0.83&0.75\\
\midrule
\multirow{2}{*}{attiro} & \multirow{2}{*}{28}  & \multirow{2}{*}{9}   & \multirow{2}{*}{59}  & \multirow{2}{*}{128} & 0.93&1.24&0.55&0.53&0.18&0.24\\
&&&&&0.67&0.71&0.52&0.68&0.28&0.31\\
\midrule
\multirow{2}{*}{\underline{\textbf{krebs}}}   & \multirow{2}{*}{8}   & \multirow{2}{*}{7}   & \multirow{2}{*}{62}  & \multirow{2}{*}{153} &
10.91&7.23&\underline{\textbf{1.19}}&0.94&\underline{\textbf{0.94}}&0.48\\
&&&&&10.39&5.56&\underline{\textbf{0.86}}&1.02&\underline{\textbf{0.57}}&0.38\\
\midrule
\multirow{2}{*}{dolphins}       & \multirow{2}{*}{20}  & \multirow{2}{*}{7}   & \multirow{2}{*}{62}  & \multirow{2}{*}{159} & 14.75&23.70&1.81&2.35&1.70&1.02 \\
&&&& & 10.66&13.83&2.80&2.98&0.74&1.50 \\
\midrule
\multirow{2}{*}{\underline{\textbf{prison}}}  & \multirow{2}{*}{28}  & \multirow{2}{*}{9}   & \multirow{2}{*}{67}  & \multirow{2}{*}{142} 
&5.90&10.22&\underline{\textbf{0.83}}&1.56&\underline{\textbf{0.62}}&0.61\\
&&&& &6.93&10.23&\underline{\textbf{4.81}}&1.03&\underline{\textbf{0.66}}&0.48\\
\midrule\newpage
\multirow{2}{*}{\underline{\textbf{huck}}}    & \multirow{2}{*}{5}   & \multirow{2}{*}{5}   & \multirow{2}{*}{69}  & \multirow{2}{*}{297}
&519.95&	299.12&	\underline{\textbf{19.53}}&	17.79&	\underline{\textbf{4.31}}& 	4.51\\
&&&& & 447.72&	493.74&	\underline{\textbf{18.22}}&	19.71	&\underline{\textbf{3.34}}&	3.31\\
\midrule
\multirow{2}{*}{sanjuansur}     & \multirow{2}{*}{35}  & \multirow{2}{*}{11}  & \multirow{2}{*}{75}  & \multirow{2}{*}{144} 
&6.16&	5.85&	0.68&	0.71&	0.37&	0.49\\
&&&& &7.70&	3.61&	0.82&	1.36&	0.44&	0.38\\
\midrule
\multirow{2}{*}{jean}   & \multirow{2}{*}{7}   & \multirow{2}{*}{5}   & \multirow{2}{*}{77}  & \multirow{2}{*}{254}
&276.98&	147.77&	15.93&	14.57&	2.45&	2.41\\
&&&& &150&147.85&13.97&14.80&2.32&2.41\\
\midrule
\multirow{2}{*}{david}  & \multirow{2}{*}{15}  & \multirow{2}{*}{8}   & \multirow{2}{*}{87}  & \multirow{2}{*}{406} 
&544.99&308.06&46.23&54.23&5.22&4.36\\
&&&& & 219.14&	323.96&	45.24&	38.33&	3.31&	3.47\\
\midrule
\multirow{2}{*}{ieeebus}        & \multirow{2}{*}{32}  & \multirow{2}{*}{13}  & \multirow{2}{*}{118} & \multirow{2}{*}{179}
&2.52&5.14&0.76&0.94&0.43&0.62\\
&&&& & 7.82&5.82&0.79&1.35&0.33&0.3\\
\midrule
\multirow{2}{*}{\underline{\textbf{sfi}}}     & \multirow{2}{*}{3}   & \multirow{2}{*}{3}   & \multirow{2}{*}{118} & \multirow{2}{*}{200} 
&6.98&6.51&\underline{\textbf{0.74}}&0.90&\underline{\textbf{0.3}}&0.31\\
&&&& & 6.40&	2.80&	\underline{\textbf{0.84}}&	1.32&	\underline{\textbf{0.34}}&	0.31\\

\midrule
\multirow{2}{*}{anna}   & \multirow{2}{*}{15}  & \multirow{2}{*}{\small\showclock{1}{00}} & \multirow{2}{*}{138} & \multirow{2}{*}{493}
&90.75&	52.11&	10.60&	23.65&	1.37&	1.71\\
&&&& & -   & -    & -    & -     & -     & -    \\
\midrule
\multirow{2}{*}{494bus} & \multirow{2}{*}{116} & \multirow{2}{*}{\small\showclock{1}{00}} & \multirow{2}{*}{494} & \multirow{2}{*}{586} 
&108.48&	126.77&	27.13&	33.09&	2.73&	2.10\\
&&&&& -   & -    & -    & -     & -     & -    \\
\midrule
\multirow{2}{*}{average}        & \multirow{2}{*}{}    & \multirow{2}{*}{}    & \multirow{2}{*}{}    & \multirow{2}{*}{}    
&84.19&	52.63&	7.02&	8.41&	1.21&	1.09\\
&&&&&51.52&	59.70&	5.75&	5.45&	0.9&	0.92\\
\bottomrule

\label{table:exper1}
\end{longtable}
\end{center}

The various methods exhibit diverse performance characteristics in terms of execution time and the number of instances solved optimally. Key observations from Table \ref{table:exper1} are as follows:

\begin{itemize}
    \item \emph{cec}2 outperforms \emph{cec} in 13 cases, \emph{ILP}$_{cut}$, \emph{ILP}$_{cut}$2, \emph{LIC} and \emph{LIC}2 in all the cases.
    \item \emph{ILP}$_{cut}$2 was faster than \emph{LIC}2 for 15 cases and \emph{LIC} for all instances.
    \item \emph{LIC}2 outperforms \emph{LIC} in 14 cases.
    \item For some instances in \emph{ILP}$_{cut}$ and \emph{cec}, the graphs and results are indicated by boldface and underlined in Table \ref{table:exper1}. This is to emphasize that these graphs contain multiple longest induced cycles of the same length, and the procedures described in Algorithm \ref{alg:CallBack1} cut the cycle if its length is less than or equal to the longest induced cycle found so far. Thus, for these graphs, all the longest cycles are found by the method. 
    \item Using \emph{LISC} as an initial solution does not contribute significantly to improving the execution time in the majority of cases.
    \item The results emphasize the correlation between graph density and execution time. Graph density is defined as the ratio of the edges present in a graph to the maximum number of edges it can hold. This relationship is particularly evident for dense graphs like huck, jean, and david, especially in the case of \emph{LIC} and \emph{LIC}2 models. However, it is not the case for \emph{cec} and \emph{cec}2 models as their running times show less sensitivity to the graph's density.
\end{itemize}

\begin{table*}[t]
\centering
\caption{Running times on MG instances, time is given in seconds.}
\begin{tabular}{c|c| c c c c c c}
\toprule
nr.~of edges & nr.~of instances & \emph{LIC} &\emph{LIC}2 & \emph{ILP$_{cut}$}& \emph{ILP$_{cut}2$}& \emph{cec}& \emph{cec}2 \\
\midrule
1--49&107&0.14&0.14&0.12&0.18&0.09&0.08\\
50--74&135&0.37&0.29&0.2&0.3&0.17&0.12\\
75--99&151&0.7&0.58&0.32&0.39&0.22&0.17\\
100--124&121&1.74&1.32&0.51&0.65&0.29&0.24\\
125--149&90&4.5&3.67&0.82&0.99&0.37&0.34\\
150--199&89&10.45&7.48&1.49&1.7&0.56&0.49\\
200--629&80&169.38&110.49&13.28&16.39&2.41&1.9\\ \hline
average&&157.23&126.35&44.73&57.28&26.84&22.01\\
\bottomrule
\end{tabular} 
\label{table:exper2}
\end{table*}

The results for the MG instances, organized into groups based on the number of edges, are presented in Table \ref{table:exper2}. Unlike \emph{LIC} and \emph{LIC}2, where the running times increase proportionally with the instance size, the results indicate that \emph{cec} and \emph{cec}2 are more reliable, with running times showing less sensitivity to the graph's size.

\begin{table}[b]
\centering
\caption{Running times on random instances for \emph{cec}2, \emph{BC}1, \emph{BC}2, and \emph{BC}3, time is given in seconds. Values in brackets show the original execution time of \emph{cec}2.}
\begin{tabular}{ccccc}
\toprule
\multicolumn{5}{c}{\underline{Randomly generated graphs: 10\% density}}\\[0.2cm]
$n$&\emph{cec}2& \emph{BC}1& \emph{BC}2& \emph{BC}3\\
\midrule
50&0.32 (0.49)&0.33&0.3&0.39\\
60&0.79 (1.2)&1.19&1.2&1.34\\
70&4.17 (6.32)&5.57&4.93&5.58\\
80&20.5 (31.1)&37.15&26.9&27.34\\
90&93.93 (142.32)&160.1&155.82&168.02\\
100&518.75 (785.99)&1321.41&1129.47&1094.8\\
\midrule
average&106.41&254.29&219.77&216.25\\
\midrule
\multicolumn{5}{c}{\underline{Randomly generated graphs: 30\% density}}\\[0.2cm]
$n$&\emph{cec}2&\emph{BC}1&\emph{BC}2&\emph{BC}3\\
\midrule
50&4.15 (6.28)&8.21&9.6&8.78\\
60&26.14 (39.6)&39.82&46.18&51.9\\
70&90.11 (136.52)&234.45&206.36&283.28\\
80&203.81 (308.8)&935.78&676.52&1139.55\\
90&544.88 (825.58)&2072.16&1874.91&3011.17\\
100&1810.49 (2743.17)&\showclock{0}{45}&\showclock{0}{45}&\showclock{0}{45}\\
\midrule
average& 446.6&	658.08&	562.71&	898.94\\
\bottomrule
\end{tabular}
\label{table:random}
\end{table}

The results for the random graphs are presented in Table \ref{table:random}, where we compare \emph{cec}2 against the top three algorithms introduced in \cite{pereira}. The runtime represents the average duration on the ten graphs in each case. Notably, \emph{cec}2 outperforms these algorithms in all cases, even before normalizing the execution times with the ratio listed in Table \ref{table:cpu}. Moreover, \emph{cec}2 successfully solved the instance with 100 vertices and 30\% density, a scenario where none of the algorithms in \cite{pereira} succeeded.

\section{Conclusion}
\label{sec:con}
Considering that the longest induced cycle problem is NP-hard, it is essential to find an efficient approach that can yield optimal solutions within a reasonable time. In this regard, we introduced three integer linear programs, some of which are extensions of models originally formulated for solving the longest induced path problem. These newly proposed programs showed differing execution times and success rates in terms of achieving optimal solutions, outperforming the models presented in the literature. 
We have found that the $cec2$ formulation with tough cut generation yields the most efficient method.

\section*{Acknowledgements}
The research leading to these results has received funding from the national project TKP2021-NVA-09. Project no. TKP2021-NVA-09 has been implemented with the support provided by the Ministry of Innovation and Technology of Hungary from the National Research, Development and Innovation Fund, financed under the TKP2021-NVA funding scheme.
The work was also supported by the grant SNN-135643 of the National Research, Development, and Innovation Office, Hungary.
\bibliographystyle{spmpsci}

\begin{thebibliography}{10}
\providecommand{\url}[1]{{#1}}
\providecommand{\urlprefix}{URL }
\expandafter\ifx\csname urlstyle\endcsname\relax
\providecommand{\doi}[1]{DOI~\discretionary{}{}{}#1}\else
\providecommand{\doi}{DOI~\discretionary{}{}{}\begingroup
\urlstyle{rm}\Url}\fi

\bibitem{pereira}
Pereira, Dilson Lucas and Lucena, Abilio and Salles da Cunha, Alexandre and Simonetti, Luidi: { Exact solution algorithms for the chordless cycle problem.}
\newblock INFORMS Journal on Computing. \textbf{34}(4),1970--1986 (2022)

\bibitem{Marzo}
Marzo, Rusl{\'a}n G and Melo, Rafael A and Ribeiro, Celso C and Santos, Marcio C: { New formulations and branch-and-cut procedures for the longest induced path problem. }
\newblock Computers \& Operations Research. \textbf{139}, 105627 (2022)

\bibitem{Bokler}
Bökler F., Chimani M., Wagner M.H., Wiedera T: { An experimental study of ILP formulations for the longest induced path problem.}
\newblock International Symposium on Combinatorial Optimization, 89--101 (2020)

\bibitem{gurobi}
Gurobi Optimization Inc. Gurobi Optimizer Reference Manual \textbf(2020)

\bibitem{Matsypura}
Matsypura, Dmytro and Veremyev, Alexander and Prokopyev, Oleg A and Pasiliao, Eduardo L: { On exact solution approaches for the longest induced path problem.}
\newblock European Journal of Operational Research \textbf{278}(2),546--562 (2019)

\bibitem{kaminski}
Kaminski, Jermain and Schober, Michael and Albaladejo, Raymond and Zastupailo, Oleksandr and Hidalgo, C{\'e}sar: { Moviegalaxies-social networks in movies.}
\newblock Harvard Dataverse, (2018)


\bibitem{kumar}
Kumar, Parveen and Gupta, Nitin: { A heuristic algorithm for longest simple cycle problem.}
\newblock Proceedings of the International Conference on Wireless Networks (ICWN), pp.~202--208 (2014)

\bibitem{broersma}
Broersma, Hajo and Fomin, Fedor V and van’t Hof, Pim and Paulusma, Dani{\"e}l: { Exact algorithms for finding longest cycles in claw-free graphs.}
\newblock Algorithmica \textbf{65}(1), 129--145 (2013)


\bibitem{walsh}
Walsh, Toby: { Symmetry breaking constraints: Recent results.}
\newblock Proceedings of the AAAI Conference on Artificial Intelligence \textbf{26}(1), 2192--2198 (2012)

\bibitem{Lokshtanov}
Lokshtanov, Daniel: { Finding the longest isometric cycle in a graph. }
\newblock Discrete Applied Mathematics.\textbf{157}(12),2670--2674 (2009)

\bibitem{chen2007}
Chen, Yijia and Flum, Jorg: { On parameterized path and chordless path problems. }
\newblock Twenty-Second Annual IEEE Conference on Computational Complexity (CCC'07), pp.~250--263 (2007)

\bibitem{fuchs}
Fuchs, Elena D: {Longest induced cycles in circulant graphs.}
\newblock The Electronic Journal of Combinatorics, Article Number R52 (2005)

\bibitem{west}
West, Douglas Brent and others: { Introduction to Graph Theory. }
\newblock Prentice Hall (2001)

\bibitem{borndorfer1998}
Bornd{\"o}rfer, Ralf: { Aspects of set packing, partitioning, and covering.}
\newblock PhD thesis, TU Berlin, Berlin. \textbf(1998)



\bibitem{nemhauser1992}
Nemhauser, George L and Sigismondi, Gabriele: { A strong cutting plane/branch-and-bound algorithm for node packing.}
\newblock Journal of the Operational Research Society. \textbf{43}(5), 443--457 (1992)

\bibitem{garey}
Garey, Michael R. and Johnson, David S.: { Computers and Intractability; A Guide to the Theory of NP-Completeness.}
\newblock W. H. Freeman \& Co. \textbf(1990)


\bibitem{wojciechowski}
Wojciechowski, Jerzy Marian: { Long induced cycles in the hypercube and colourings of graphs.}
\newblock University of Cambridge (1990)


\bibitem{alon}
Alon, Noga: { The longest cycle of a graph with a large minimal degree.}
\newblock Journal of Graph Theory. \textbf{10}(1), 123--127 (1986)

\bibitem{akiyama}
Akiyama, Takanori and Nishizeki, Takao and Saito, Nobuji: { NP-completeness of the Hamiltonian cycle problem for bipartite graphs.}
\newblock Journal of Information Processing. \textbf{3}(2), 73--76 (1980)



\bibitem{cpu}
PassMark Software - CPU Benchmarks: {\url{ https://www.cpubenchmark.net/compare}} \textbf(2023)
\end{thebibliography}

\end{document}